# Experience of Developing a Meta-Semantic Search Engine

Debajyoti Mukhopadhyay[1], Manoj Sharma[1], Gajanan Joshi[1], Trupti Pagare[1], Adarsha Palwe[1]
Department of Information Technnology[1]
Maharashtra Institute of Technology
Pune 411038, India
{debajyoti.mukhopadhyay, manojsharma2708, joshi.gajanan09,
pagaretrupti2, adarsha125}@gmail.com

*Abstract*— Thinking of today's web search scenario which is mainly keyword based, leads to the need of effective and meaningful search provided by Semantic Web. Existing search engines are vulnerable to provide relevant answers to users query due to their dependency on simple data available in web pages. On other hand, semantic search engines provide efficient and relevant results as the semantic web manages information with well defined meaning using ontology. A Meta-Search engine is a search tool that forwards users query to several existing search engines and provides combined results by using their own page ranking algorithm. SemanTelli is a meta-semantic search engine that fetches results from different semantic search engines such as Hakia, DuckDuckGo, SenseBot through intelligent agents. This paper proposes enhancement of SemanTelli with improved snippet analysis based page ranking algorithm and support for image and news search.

## I. Introduction

Search Engines has become one of the most important and used tool over the World Wide Web. Search Engines normally search web pages for the required information and then display the results by using ranking algorithms. These commercially available search engines do not completely serve the needs and demands of the users. Most of the times the relevancy of the provided results is not accurate. These types of problem occur because of the structure of the current Web.

**Intelligent Agent**

Intelligent Agents are an emerging technology that is making computer systems easier to use by allowing people to delegate work back to the computer. They help to do things like finding and filtering information, customize views of information and automate work. An intelligent agent is software that assists people and acts on their behalf. Agents can intelligently summarize data, learn from us and even can make recommendations to us.

- Delegation aspect of intelligent agents, which is certainly one thing that sets agents apart, since agents are built to help people, just like human assistants.
- All agents are autonomous that is an agent has control over its own actions.
- All agents are goal driven. Agents have a purpose and act in accordance with that purpose.

**Semantic Search Engines**

Semantic Search Engines makes use of Semantic Web in order to search relevant results. These SE's manage the information over web using ontology, developed with the help of OWL (Web Ontology Language). Ontology arranges the pieces of information in a hierarchical tree like structure, where information sharing some properties is grouped together. Searching is carried out by traversing the tree with efficient algorithms. These techniques make semantic search faster and relevant. Here, details of the semantic search engines used in SemanTelli are provided.

- **DuckDuckGo**

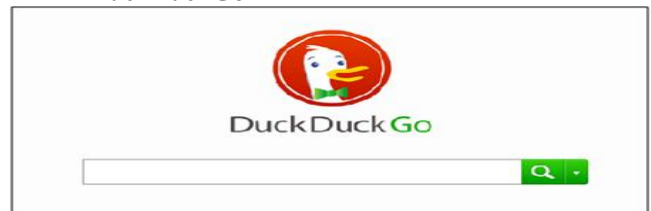

**Fig. 1.** DuckDuckGo semantic search engine

The main sources of information for DuckDuckGo are Crowd-sourced websites that makes it capable to enhance traditional results and improve their relevance. It provides policy that values privacy and does not record user information as users are not profiled.

- **Hakia**

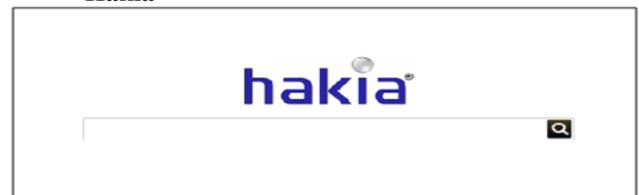

**Fig. 2.** Hakia semantic search engine

Hakia uses enhanced semantic technology for ranking algorithm that uses QDEX(Query Detection and Extraction). It perceives quality results from all segments like news, blogs, credible, hakia galleries, images and videos. For long queries, it highlights relevant phrases, keywords or sentences that relates to users query.

- **Sensebot**

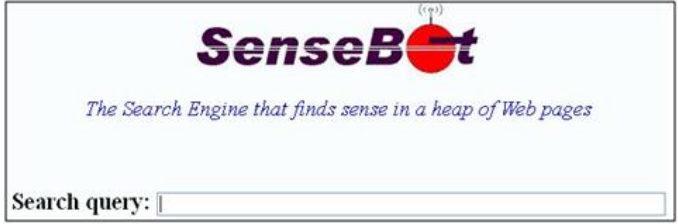

**Fig. 3.** Sensebot semantic search engine

Sensebot prepares the text summary according to the user's search query. It identifies key semantic concepts by using text mining algorithms that parse the Web Pages. The retrieved multiple documents are then used to perform a coherent summary. This coherent summary becomes the final result for user's query. The main sources for these results are usually the news agencies.

## II. SemanTelli Architecture

SemanTelli is a simple meta-semantic-search engine which extends the concept of meta-search engines to enhance searching of specific data semantically. Figure 4 shows the whole architecture of SemanTelli with its major functioning blocks.

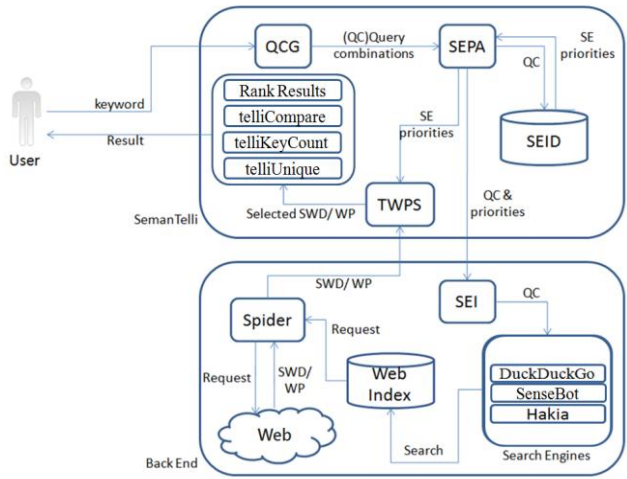

**Fig. 4.** Architecture of SemanTelli

Abbreviations used:
- QCG : Query Combination Generator
- SEPA : Search Engine Priority Assigner
- SEI : Search Engine Interactor
- TWPS : Temporary Web Page Storage
- SEID : Search Engine Information Database

Our proposed architecture is comprised of two blocks, as given below

**SemanTelli**

SemanTelli comprises of user interface and other modules to find out search engines (SE) priorities and to filter the obtained search results from search engines. Query Combination Generator (QCG) accepts the query from user and generates its possible and meaningful combinations considering proper sequences of query keywords. These combinations are sent to SEPA (Search Engine Priority Assigner) which finds the priority for different SE's to monitor search results. SEPA uses Search Engines Information Database (SEID) which contains information about importance of each search engine for particular keyword or its domain. This information helps SEPA to give specific priorities to each SE. Here, for now, we have assigned priorities to different search engines on basis of a survey with sample set of queries and expected results. SEPA sends these priorities and QC's to SE interface (SEI). The search results consisting of Semantic Web Documents (SWD) and Web Pages (WP) are returned by backend to Temporary Web Page Storage (TWPS), which stores the results temporary. TEMPS uses the properties from SEPA to filter out the SWD/WP's and sends the required results for further refinements.

Here, many selected SWD/WP's have to go through our algorithm in order to reduce redundancy and provide actual needed output to the user. Our approach focuses on analysis of snippets obtained in results returned by search engines. This snippet analysis consists of extraction of keywords from snippet, comparing their count and sequence with query keywords. Finally, results are arranged in decreasing order of relevance considering their keyword count, keyword sequence and initial weight ($W_i$).

**Back End**

This block gets the QC and SE priorities from SEPA and searches the SWD's and WP's using search engines index over the World Wide Web through our different semantic search engines. Here, spiders are located over the web and make searching of web pages on web easier. Spiders also send these web pages to TWPS for further processing and refinement.

## III. CURRENT STATUS OF SEMANTELLI

SemanTelli is in developing phase. It fetches the result from semantic search engines, analyzes snippet of each result and arranges these results. We have assigned initial weight to each fetched result as mentioned in Table. 1.

| Sr. No. | Search Engine | Initial Weight ($W_i$) |
|---|---|---|
| 1 | DuckDuckGo | 0.3 |
| 2 | Hakia | 0.2 |
| 3 | SenseBot | 0.1 |

**Table 1.** Initial Weight

Considering relevancy of results, we have assigned more initial weight to DuckDuckGo results as compared to Hakia and SenseBot results.

Figure 5 highlights the implementation details of SemanTelli. It shows that fetched results are temporarily stored in respective buffers. These results are traversed in order to analyze snippets which indicate their relevance. The relevance of results is found out on basis of keyword count within snippet, keyword sequence with respect to query keywords and assigned initial weight ($W_i$). After post processing, results are arranged with respect to their relevance.

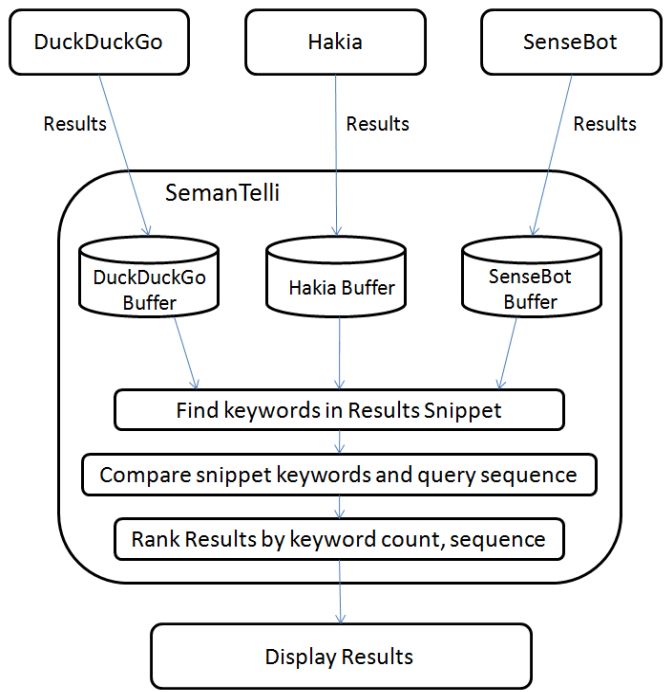

**Fig. 5.** Implementation details

Following algorithm is used to find out relevance of results.

```
Algorithm: Ranking Algorithm using
           Snippet Analysis
Input:     Obtained results
Output:    Ranked results

Step1: Apply Conflation algorithm on
       query.
       a) Removal of stop words (most
          common occurring words).
       b) Suffix Stripping (remove the
          suffixes of words and retrieve
          stems (keyword). E.g. moving ->
          move).
       c) Retrieve the keywords.

Step2: Apply similar conflation algorithm
       as in step 1 to all obtained
       result snippets.

Step3: Count the number of keywords from
       snippet and query.
       Consider,
         NK_s = no of keywords in snippet.
         NK_Q = no of keywords in query.
Step4: For each keyword K_S in snippet
         count(K_S) = 0
         For each keyword K_Q in query
             If K_S and K_Q are same
                 count(K_S) ++
Step5: Arrange results on the basis of
       decreasing order of keyword count.
Step6: If keyword count (count(K_S)) of
       more than one snippets is
       identical, then
       a) Check the sequence of
          keywords in snippet with
          sequence in query.
       b) Specific snippet having
          similar sequence as in
          query will get more
          priority.
Step7: If sequence of keywords among more
       than one snippet is identical,
       then
         Arrange the selected results
       considering their ranking from
       respective search engine and
       considered initial weight (W_i).
Step8: Return the ranked results.
```

Using snippet analysis we are getting only relevant results and those results are presented to the user in decreasing order of relevance.

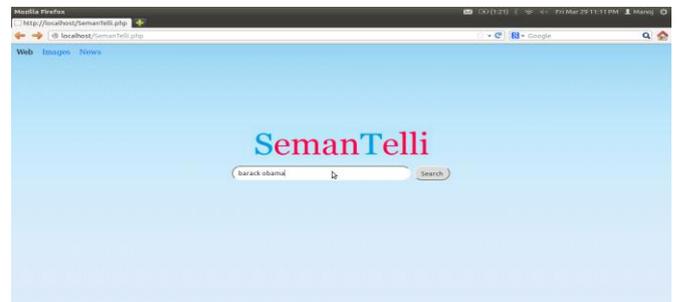

**Fig. 6.** Home page of SemanTelli.

**Fig 6** shows the homepage of SemanTelli. In this figure user has submitted Barack Obama as a query. This is query is sent to the semantic search engines and results are fetched. After fetching results the algorithm is applied and results are presented to the user.

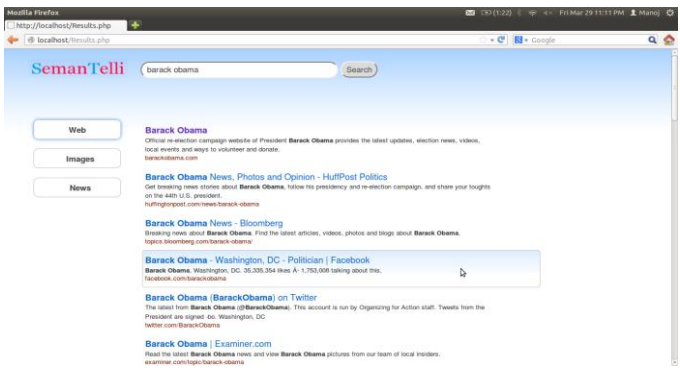

**Fig. 7.** Web results of SemanTelli

**Fig 7** shows web results of SemanTelli. The query is Barack Obama for Web results. The results are fetched from semantic search engines, combined and presented to the user.

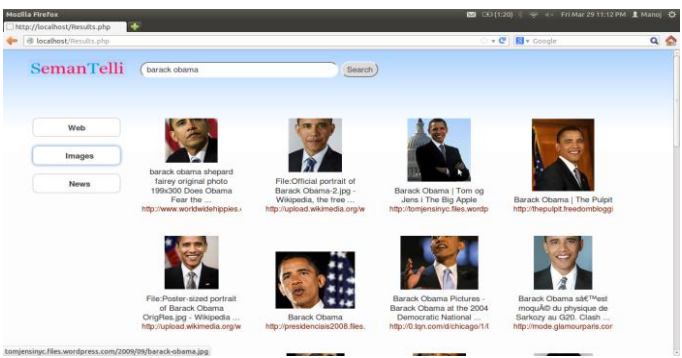

**Fig. 8.** Image results of SemanTelli.

**Fig 8** shows image results of SemanTelli. The query is same as Barack Obama and SemanTelli provides only relevant images.

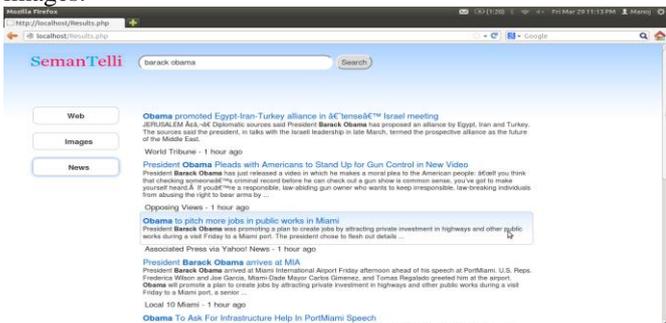

**Fig. 9.** News results of SemanTelli.

**Fig 9** shows news results of SemanTelli. The news results are sorted in descending order of time i.e. the most recent news is shown first.

## IV. PERFORMANCE ANALYSIS

Current development of SemanTelli is based on Snippet Analysis of obtained results. Considering many sample queries, the ranking of results is well improved and arranges the results in more relevant fashion. Snippet Analysis provided an efficient way to dig out relevance of results as per query keywords. Hence, this ranking approach is unique as compared to other approaches which considers hit counts, last modification date, etc. However, our approach is still in development phase for further refinement of ranking algorithm to get relevant results. Still, there are some aspects considering response time and fast post-processing of results that can be improved.

## V. CONCLUSION

We have developed a meta-semantic-search engine "SemanTelli" which highlights the novel concept of integrating search results from existing semantic search engines such as DuckDuckGo, Hakia, SenseBot and rank them with unique snippet analysis based approach. This will provide users with an efficient way to search for required results from different semantic search engines by single interface. SemanTelli mines deeply through the combinations of user query using semantic search engines and brings out only relevant results with help of snippet analysis.